\documentclass[12pt]{article}

\usepackage{amsmath, amsthm, amsfonts, amssymb, bm, bbm, xcolor}
\usepackage{algorithm, algorithmic, graphicx, subcaption}
\usepackage{hyperref, xcolor, tabularx, multirow, booktabs, enumerate, natbib, xr}

\newtheorem{Theorem}{Theorem}

\newtheorem{Example}{Example}

\newtheorem{Proposition}{Proposition}

\newcommand{\indep}{\;\, \rule[0em]{.03em}{.67em} \hspace{-.27em} \rule[-.02em]{.7em}{.03em} \hspace{-.27em} \rule[0em]{.03em}{.67em}\;\,}

\newcommand{\blind}{1}

\begin{document}

\if1\blind
{
\title{\Large{\textbf{Robust Learning of Heterogeneous Dynamic Systems}}} 
\author{
\bigskip
\large{Shuoxun Xu$^{1}$, Zijian Guo$^{2}$, Brooke R. Staveland$^{3}$,} \\
\bigskip
\large{Robert T. Knight$^{1}$, and Lexin Li$^{1}$\thanks{Corresponding author}} \\
\normalsize{\textit{$^1$University of California at Berkeley}} \\
\normalsize{\textit{$^2$ Zhejiang University}} \\
\normalsize{\textit{$^3$University of California at San Francisco}} 
}
\date{}
\maketitle
} \fi

\if0\blind
{
\title{\Large{\textbf{Robust Learning of Heterogeneous Dynamic Systems}}}
\author{
\bigskip
\vspace{1.35in}
}
\date{}
\maketitle
} \fi

\baselineskip=20pt
\begin{abstract}
Ordinary differential equations (ODEs) provide a powerful framework for modeling dynamic systems arising in a wide range of scientific domains. However, most existing ODE methods focus on a single system, and do not adequately address the problem of learning shared patterns from multiple heterogeneous dynamic systems. In this article, we propose a novel distributionally robust learning approach for modeling heterogeneous ODE systems. Specifically, we  construct a robust dynamic system by maximizing a worst-case reward over an uncertainty class formed by convex combinations of the derivatives of trajectories. We show the resulting estimator admits an explicit weighted average representation, where the weights are obtained from a quadratic optimization that balances information across multiple data sources. We further develop a bi-level stabilization procedure to address potential instability in estimation. We establish rigorous theoretical guarantees for the proposed method, including consistency of the stabilized weights, error bound for robust trajectory estimation, and asymptotical validity of pointwise confidence interval. We demonstrate that the proposed method considerably improves the generalization performance compared to the alternative solutions through both extensive simulations and the analysis of an intracranial electroencephalogram data.
\end{abstract}


\newpage
\baselineskip=22pt

\section{Introduction}
\label{sec:introduction}

Ordinary differential equations (ODEs) provide a mathematical foundation for modeling dynamic systems, by relating the rate of change of variables to their current states, thereby  describing how a system evolves over time. ODEs have been widely used across a range of scientific disciplines, for instance,  infectious disease \citep{wu2005statistical, Liang2008}, computational biology \citep{CaoZhao2008, Chou2009, ma2009defining}, and neuroscience \citep{Izhikevich2007, Zhang2017, CaoLuo2019}. There have been numerous ODE models developed in recent years, including linear ODEs \citep{lu2011high, zhang2015dynamic, dattner2015optimal}, additive ODEs \citep{henderson2014network, wu2014sparse, chen2017network},  kernel ODEs \citep{dai2022kernel, DaiLi2024postinf}, and neural ODEs \citep{chen2018neural}, among others. Despite their success, most existing ODE methods primarily focus on modeling a \emph{single} dynamic system. Approaches for handling \emph{multiple}, especially \emph{heterogeneous}, dynamic systems remain largely unexplored.

Our motivation arises from an approach-avoidance conflict (AAC) study that employs intracranial electroencephalogram (iEEG) to investigate prefrontal-hippocampal brain networks \citep{staveland2025circuit}. Approach-avoidance conflict describes daily situations where a single choice entails potential gains and losses simultaneously, which induces anxiety in everyday life. Understanding brain networks that regulate AAC is crucial to understanding the neural underpinnings of both healthy and pathological anxiety. Intracranial EEG is a neurophysiological technique in which electrodes are placed directly on the surface of the brain to record electrical activity with high spatial and temporal resolutions. In this study, five epilepsy patients were recruited to perform a continuous-choice, AAC decision-making task inspired by the arcade game Pac-Man, while their brain activities were recorded from six electrodes implanted in six different brain regions. A key scientific objective is to uncover brain connectivity patterns from the iEEG signals during the AAC task. Whereas ODEs provide a suitable framework for modeling such brain dynamics for each individual patient, there are some major challenges that complicate their application for multiple patients. There is substantial patient-to-patient variability, which reflects inherent differences in individual brain anatomy and function. Besides, the number of patients is extremely limited, which is common to iEEG studies due to the highly invasive nature of the iEEG technology. A crucial scientific question is therefore how to obtain a reliable estimate of the underlying brain connectivity network by appropriately combining information from multiple patients. However, how to aggregate such information in a principled way is far from straightforward, and remains a major methodological challenge, one that is not unique to this particular example but is commonly encountered across many scientific applications. 

There have been various strategies developed to integrate knowledge from multiple patients, or sources, which can be grouped into several categories. The first category builds on empirical risk minimization, which learns a predictive model by pooling together multiple data sources and minimizing the average risk over the combined data \citep{bazeley2012integrative}. In the context of ODE modeling, \citet{chen2017network, dai2022kernel} employed this strategy by minimizing the sum of empirical losses across multiple experiments. While simple and effective in many applications, this approach typically assumes that all data sources share a common underlying model, and its performance degrades in the presence of substantial heterogeneity among sources. The second category relies on the notion of invariance, which learns a stable predictive model by optimizing an invariance score and identifying a model that is invariant across heterogeneous sources \citep{peters2016causal, pfister2019learning}. In ODE modeling, \citet{dai2022kernel} adopted this strategy and proposed an invariance score along with their kernel ODEs. This approach explicitly accounts for data heterogeneity, but may exhibit limited generalizability when the number of data sources is small. The third category involves federated regression and predictive learning, which trains a global model across multiple distributed data sources without transferring raw data \citep{LianFan2018, HanCai2025, ChenTang2025}. However, their primary focus is on preserving data privacy, and none directly targets ODE, whereas our goal centers on addressing data heterogeneity with a limited number of sources in ODE modeling.

There is another category of approaches, known as \emph{distributionally robust learning}, which aims to build a robust and stable predictive model across multiple sources \citep[][among others]{hu2018does, sagawa2019distributionally, zhang2020coping, LinFangGao2022, guo2023statistical, wang2023distributionally, GuoLiHanCai2025}. The key idea is to construct a model whose performance remains reliable not only on the observed training distribution, but also under the most adverse distributional shifts within a prespecified class of plausible alternatives. By explicitly optimizing against the worst-case loss over this uncertainty class, it discourages reliance on any single data source or distributional realization, thereby yielding a solution that is more stable and generalizable in the presence of data heterogeneity. Nevertheless, applying this principle to dynamic system modeling requires substantial, highly nontrivial extensions.

In this article, we propose a novel ODE modeling approach for uncovering scientific patterns that are generalizable and stable across multiple heterogeneous dynamic systems. Our approach is grounded in distributionally robust learning, which provides a principled way to handle heterogeneity and improve generalization beyond any single data source. Specifically, we construct a robust dynamic system by optimizing a worst-case objective over a targeted class of ODE trajectories, which are characterized through convex combinations of the derivatives of trajectories from multiple heterogeneous sources. We show that the resulting solution admits the form of a weighted average of the observed trajectories, whereas the weights are obtained by solving a quadratic optimization that leverages the observed signal trajectories to properly balance the influence of each individual source.  To address potential non-uniqueness and instability in the adversarial optimization, we further introduce a bi-level procedure that yields a stable and well-defined population target. We establish rigorous theoretical guarantees for the proposed method, including robustness and consistency properties. We validate our approach through extensive simulations and the motivating iEEG-based AAC study, demonstrating its clear empirical advantages over alternative methods based on empirical risk minimization and invariance estimation.

Meanwhile, our method differs in several important ways from existing distributionally robust learning solutions that primarily focus on regression settings. First, we construct the uncertainty class as all possible convex combinations of the observed dynamic systems, defined through their derivative functions rather than their link functions. This differs from  regressions, as in ODE models the derivatives are inherently intertwined with the system trajectories, which compromises convexity and complicates subsequent optimization. Second, we define the reward function used to evaluate models within the uncertainty class based on the derivative functions too. This again departs from regressions in which the reward is constructed from fully observed responses, whereas the derivative functions in our setting are latent. Their estimation therefore requires smoothing or numerical differentiation, introducing additional errors and complicating both model fitting and inference. Finally, the robust learning can be highly unstable for ODE modeling when some key matrix is or nearly singular. Ridge regularization, as currently used in \citet{guo2023statistical}, cannot fully resolve this issue. We propose a bi-level optimization strategy, in which the first level identifies all optimal solutions and the second level selects the one with the smallest norm to ensure stability. In summary, we extend distributionally robust learning to the ODE setting, broadening its scope beyond predictive regression to dynamic system modeling. Such an extension is far from trivial, and helps address an important class of scientific questions.

We adopt the following notation throughout this article. We use $D_t f(t)$ to denote the derivative function of $f(t)$, i.e., $D_t f(t) = d f(t) / dt$. For a positive integer $n$, let $[n] = \{1, \ldots, n\}$. For real numbers $a$ and $b$, let $a \vee b = \max\{a, b\}$ and $a \wedge b = \min\{a, b\}$. For two sequences $a_n$ and $b_n$, we write $a_n \lesssim b_n$ if there exists a universal constant $C > 0$ such that $a_n \leq C b_n$ for all sufficiently large $n$. For a vector $x \in \mathbb{R}^p$, let $\|x\|_q = (\sum_{j=1}^p |x_j|^q)^{1/q}$ denote its $\ell_q$-norm, and when $q=2$, $\|x\|_2$ is the Euclidean norm. For a matrix $A$, let $\|A\|_F$ denote the Frobenius norm, $\|A\|_2$ the spectral norm, and $\|A\|_\infty$ the elementwise maximum norm. For a real-valued function $f$, let $\|f\|_\infty = \sup_x |f(x)|$. For a sequence of random variables $X_n$, we write $X_n = O_p(a_n)$ to indicate that $X_n/a_n$ is bounded in probability. Unless otherwise specified, all vector inequalities are elementwise.

The rest of the article is organized as follows. Section \ref{sec:model} introduces the ODE model and our proposed ODE robust learning. Section \ref{sec:procedure} develops the corresponding estimation and inference procedure. Section \ref{sec:theory} establishes the theoretical properties. Section \ref{sec:simulations} presents the simulations, and Section \ref{sec:realdata} revisits the motivating example of the AAC study. The Supplementary Appendix collects all technical proofs.

\section{ODE Model and Robust Learning}
\label{sec:model}

\subsection{ODE model setup}

Consider $K$ independent subjects or data sources, each with the $p$-dimensional signal trajectories, $Y^{(k)}(t) = (Y_1^{(k)}(t), \ldots, Y_p^{(k)}(t))^\top \in \mathbb{R}^p$, observed at $n$ time points, $t = t_1 = 0, \ldots, t_{n} \in [0,T]$, $k \in [K]$. Suppose the observed data trajectories follow the ODE model, 
\begin{align} \label{eqn:model}
\begin{split}
& Y^{(k)}(t) =  X^{(k)}(t) + \epsilon^{(k)}(t), \\
& D_t X^{(k)}(t) \equiv 
\frac{d X^{(k)}(t)}{dt} =  \left(
\begin{array}{c}
\dfrac{d X^{(k)}_1(t)}{d t} \\
\vdots\\
\dfrac{d X^{(k)}_p(t)}{d t} \\
\end{array}
\right) = \left(
\begin{array}{c}
F^{(k)}_1(X^{(k)}(t))  \\
\vdots\\
F^{(k)}_p(X^{(k)}(t)) \\
\end{array}
\right) 
\equiv F^{(k)}\left( X^{(k)}(t), t \right), 
\end{split}
\end{align}
where $X^{(k)}(t) = (X_1^{(k)}(t), \ldots,X_p^{(k)}(t))^\top \in \mathbb{R}^p$ is the vector of latent signals, $\epsilon^{(k)}(t)$ is the measurement error with mean zero, $\epsilon^{(k)}(t) \indep X^{(k)}(t)$, $\epsilon^{(k)}(t) \indep \epsilon^{(k)}(t')$ if $t\neq t'$, and $F^{(k)} = \{F_1^{(k)}, \ldots, F_p^{(k)}\}$ is the set of unknown link functions that characterize the relations among $X^{(k)}(t)$. It is important to note that, unlike the empirical risk minimization approach that imposes a \emph{common} latent signal across all $K$ subjects, we allow each subject to follow its own latent trajectory $X^{(k)}(t)$. Correspondingly, the link function $F^{(k)}$, induced through $X^{(k)}(t)$ following $D_t X^{(k)}(t) = F^{(k)}(X^{(k)}(t),t)$, is also permitted to \emph{differ} across subjects.

\subsection{Uncertainty class and reward function}

Our goal is to learn patterns that generalize across heterogeneous subjects rather than fitting a single subject's dynamics. Toward that end, we adopt the distributionally robust learning principle and introduce two key components: an uncertainty class that captures plausible variability across the observed dynamics, and a reward function that evaluates a candidate dynamic system within this class. We then aim to maximize the worst-case reward over the uncertainty class, thereby ensuring robust and reliable performance even under the most adverse admissible dynamics.

We first define the uncertainty class. Since an ODE system is essentially characterized by the latent signal trajectory, we define the uncertainty class as the one generated by all possible convex combinations of $\{ X(0), D_t X^{(k)}(t) \}, k \in [K]$, i.e., 
\begin{align} \label{eqn:uncertainty-class}
\mathcal{C} = \left\{ S = (X(0), D_t X(t)) \ \big | \ D_t X(t) = \sum_{k=1}^K \omega_k D_t X^{(k)}(t), \; \omega_k \in [0, 1], \ \sum_{k=1}^K \omega_k = 1 \right\}.
\end{align}
Here $X(0)$ denotes the initial state, and for simplicity, we assume all subjects share the same $X(0)$. It is worth noting that, the uncertainty class in \eqref{eqn:uncertainty-class} is built on the derivative space of $D_t X(t)$, rather than the link function space of $F(\cdot,t)$. In ODE systems, trajectories and their derivatives are coupled through the nonlinear dynamic functions, $D_t X(t) = F^{(k)}(X(t),t)$. For any convex set of link functions $\mathcal{F}$, when $F^{(1)}$, $F^{(2)} \in \mathcal{F}$, then $\omega F^{(1)} + (1-\omega) F^{(2)} \in \mathcal{F}$, for any $\omega \in [0,1]$. However, the induced trajectory set of $\mathcal{F}$, i.e., $\{ X(t): D_t X(t) = F(X(t),t), F\in\mathcal{F} \}$, is \emph{not} necessarily convex, because $D_t X^{(1)}(t) = F^{(1)}(X^{(1)}(t) , t)$ and $D_t X^{(2)}(t) = F^{(2)}(X^{(2)}(t) , t)$ cannot guarantee that $D_t(\omega X^{(1)}(t) + (1-\omega)X^{(2)}(t)) = (\omega F^{(1)} + (1-\omega) F^{(2)}) \circ (\omega X^{(1)}(t) + (1-\omega)X^{(2)}(t),t)$ is convex, where $\circ$ denotes the functional composition. As a result, if we were to follow the usual practice in distributionally robust regressions and define the uncertainty class in terms of the link functions, the induced set of trajectories could be non-convex, which would complicate subsequent optimization and inference. 

We next define the reward function. For any $S = \{ X(0),D_t X(t) \} \in \mathcal{C}$, define the reward function of $F = (F_1, \ldots, F_p)^\top$ as, 
\begin{align} \label{eqn:reward}
R_{S}(F) = \int_{0}^T \sum_{j=1}^p \left[ (D_t X_j(t))^2 - \{ D_t X_j(t) - F_j(X(t),t) \}^2 \right] dt. 
\end{align}
At a high level, this reward measures how much total variation of the target ODE system generated by $S$ can be explained by $F$, and a larger reward implies a better predictive performance. It is noted that, this definition is conceptually related to the reward construction in \cite{wang2023distributionally}, but differs in a crucial way.  \cite{wang2023distributionally} defined rewards directly from the observed response, whereas in our setting, both the state $X(t)$ and its derivative $D_t X(t)$ are latent. Their estimation requires smoothing or numerical differentiation, therefore introducing additional errors and complicating both model fitting and inference.

\subsection{Robust learning of ODEs}

Adopting the distributionally robust learning principle, under the given uncertainty class $\mathcal{C}$ and the reward function $R_{S}(F)$, we seek the robust link function estimator $F^*$ that maximizes the worst-case reward, i.e., the reward with the minimum explained total variation,
\begin{align} \label{eqn:adv-optim}
F^* = \arg\max_{F} \min_{S \in \mathcal{C}} R_{S}(F).
\end{align}
Such an adversarial optimization is to avoid overfitting to any single data source, and thus enhances generalization and resilience to data heterogeneity. 

The next theorem characterizes the solution to \eqref{eqn:adv-optim} and the corresponding trajectory. 

\begin{Theorem}\label{thm1}
The solution $F^*$ to \eqref{eqn:adv-optim} and the corresponding trajectory $X^*(t)$ satisfy that
\begin{align}
& F^*(X^*(t), t) = \sum_{k=1}^K \omega_k^* \ F^{(k)}(X^{(k)}(t),t) = D_t( X^*(t) ), \label{eqn:F_star} \\
& X^*(t) = \sum_{k=1}^K \omega_k^* X^{(k)}(t), \quad t \in [0, T], \label{eqn:X_star}
\end{align}
where $\omega^* = \arg\min_{\omega \in \mathcal{H}} \, \omega^{\top} \Gamma \omega$, $\mathcal{H}$ denotes the $K$-dimensional simplex with all elements being non-negative and summing up to one, $\Gamma \in \mathbb{R}^{K \times K} = (\Gamma_{k,k'})$, $k, k' \in [K]$, and 
\begin{align*}
\Gamma_{k,k'} = \sum_{j=1}^p  \int_0^T \left(F_j^{(k)}(X^{(k)}(t),t) \times F_j^{(k')}(X^{(k')}(t),t)\right) dt 
= \sum_{j=1}^p  \int_0^T \left( D_t X_j^{(k)}(t) \times D_t X_j^{(k')}(t) \right) dt.
\end{align*}
\end{Theorem}

Theorem \ref{thm1} provides an explicit characterization of the optimal distributionally robust ODE system under the chosen uncertainty class \eqref{eqn:uncertainty-class} and the reward \eqref{eqn:reward}, where the robust link function $F^*$ and the robust signal trajectory $X^*(t)$ can both be expressed as a weighted combination of the link functions $F^{(k)}$ and the latent trajectories $X^{(k)}(t)$ from the observed subjects or data sources. The optimal weights $\omega^* = (\omega_1^*, \ldots, \omega_K^*)$ are obtained by minimizing a quadratic form based on the observed trajectories, which, equivalently, maximizes the model's robustness to worst-case variability among the source systems. This weighting strategy ensures that the solution hedges against the most adverse scenario represented by convex combinations of the observed derivatives.

\section{Estimation and Inference}
\label{sec:procedure}

\subsection{Estimation with stability}

We first address the estimation, and then inference, for the robust ODE system. Theorem \ref{thm1} suggests a practical way for estimating $X^*(t)$; i.e., we first estimate each individual latent trajectory $X^{(k)}(t)$, and then estimate the optimal weights $\omega^*$. Subsequently, we learn $F^*$ by fitting an ODE model to the resulting trajectory $X^*(t)$. 

We estimate the individual trajectory $X^{(k)}(t)$ using the usual local polynomial regression or kernel smoothing; see, e.g., \citet{Varah1982, dai2022kernel}.

We estimate the weights $\omega^*$ by solving the quadratic optimization, 
\begin{align*}
\widehat{\omega}_{\text{plug-in}} = \arg\min_{\omega \in \mathcal{H}}\, \omega^\top \widehat{\Gamma} \omega,
\end{align*}
where the matrix $\widehat{\Gamma}$ summarizes the similarity between the estimated trajectories $\widehat{X}_j^{(k)}(t)$, and its entries are given by
\begin{align*}
\widehat{\Gamma}_{k,k'} = \sum_{j=1}^p \int_0^T \left[ D_t \widehat{X}_j^{(k)}(t) \times D_t \widehat{X}_j^{(k')}(t) \right] dt.
\end{align*}
However, such a naive plug-in estimator can be highly unstable if the matrix $\Gamma$ is singular or nearly singular; that is, when its minimum eigenvalue $\lambda_{\min}(\Gamma)$ is close to zero or exactly zero \citep[][Lemma 6]{guo2023statistical}. Actually, its estimation error is bounded by
\begin{align*} 
\left\| \widehat{\omega}_{\text{plug-in}} - \omega^* \right\| \lesssim \frac{\| \widehat{\Gamma} - \Gamma \|_F}{\lambda_{\min}(\Gamma)},
\end{align*}
which can become extremely large as $\lambda_{\min}(\Gamma) \to 0$. A solution commonly used in the existing distributionally robust learning literature is to add a ridge penalty term \citep{guo2023statistical}; i.e., 
\begin{align*}
\widehat{\omega}_{\text{ridge}} (\lambda) = \arg\min_{\omega \in\mathcal{H}} \omega^\top \widehat{\Gamma} \omega + \lambda \| \omega \|_2^2.
\end{align*}
Nevertheless, even with the ridge regularization, the estimator remains unstable when $\Gamma$ is singular. This can be seen from the estimation error bound, 
\begin{align*}
\|\widehat{\omega}_{\text{ridge}} (\lambda) - \omega^*\|_2 \lesssim \frac{\|\widehat{\Gamma} - \Gamma\|_F}{\lambda_{\min}(\Gamma) + \lambda} + \frac{\lambda \sqrt{K}}{\lambda_{\min}(\Gamma)},
\end{align*}
which still diverges as $\lambda_{\min}(\Gamma) \to 0$. Similarly, the difference between the ridge solutions under different regularization parameters is 
\begin{align*}
\|\widehat{\omega}_{\text{ridge}} (\lambda) - \widehat{\omega}_{\text{ridge}}(\lambda/2)\|_2 \lesssim \frac{\lambda \sqrt{K}/2}{\lambda_{\min}(\Gamma) + \lambda/2},
\end{align*}
which also diverges. These bounds suggest that neither the plug-in estimator nor the ridge regularization can provide a stable estimate of $\omega^*$ when $\Gamma$ is singular or nearly singular. 

To address this issue, we propose a bi-level optimization strategy. Specifically, when $\lambda_{\min}(\Gamma) = 0$, the solution set $\left\{\omega \in \mathcal{H} : \omega^\top \Gamma \omega = U_\Gamma\right\}$, where $U_\Gamma = \min_{\omega \in \mathcal{H}} \omega^\top \Gamma \omega$, is not a singleton, leading to ambiguity in its solution. Therefore, we propose to select, among all minimizers, the one with the minimum $\ell_2$ norm; i.e., 
\begin{align} \label{eqn:stable-wts}
\omega_{\text{stable}}^* = \arg\min_{\omega \in \mathcal{H},\, \omega^\top \Gamma \omega \leq U_\Gamma} \|\omega\|_2^2. 
\end{align}
This way, it is ensured that $\omega_{\text{stable}}^*$ is uniquely defined, it still maximizes the worst-case reward when $\Gamma$ is singular, and it equals $\omega^*$ when $\Gamma$ is non-singular. Let $U_{\widehat{\Gamma}} = \min_{\omega \in \mathcal{H}} \omega^\top \widehat{\Gamma} \omega$. We further introduce a tolerance value $d_n > 0$ to account for estimation error. Correspondingly, we seek the stabilized estimator of the weights $\omega^*$ as, 
\begin{align*}
\widehat{\omega}_{\text{stable}} = \arg\min_{\omega \in \mathcal{H},\, \omega^\top \widehat{\Gamma} \omega \leq U_{\widehat{\Gamma}} + d_n} \|\omega\|_2^2.
\end{align*}
As later shown in Section \ref{sec:theory}, with a suitable choice of $d_n$, we can ensure that, 
\begin{align*}
\left\{\omega \in \mathcal{H} : \omega^\top \Gamma \omega \leq U_\Gamma\right\} \subseteq \left\{\omega \in \mathcal{H} : \omega^\top \widehat{\Gamma} \omega \leq U_{\widehat{\Gamma}} + d_n\right\},
\end{align*}
with a high probability, provided that $P\left(d_n \geq 2 \max \big\{ |\lambda_{\min}(\widehat{\Gamma} - \Gamma)|,\, |\lambda_{\max}(\widehat{\Gamma} - \Gamma)| \big\} \right)$ approaches one when $n$ approaches $\infty$. 

\begin{figure}[t!]
\centering
\includegraphics[width=0.45\linewidth, height=2.25in]{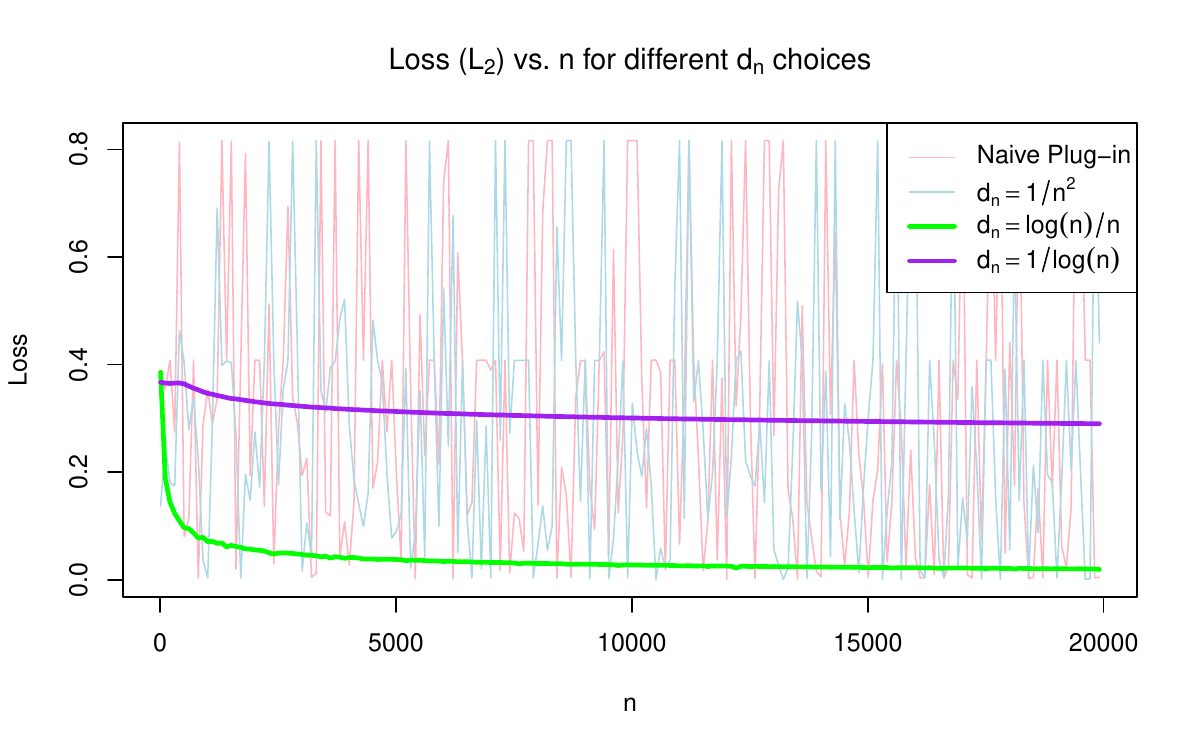}
\caption{Stability of estimating the weight $\omega^*$.}
\label{diff_dn_dist}
\end{figure}

Combining the above estimation steps, we obtain the robust estimator of the trajectory, 
\begin{align}\label{eqn:robust-est}
\widehat{X}_{\mathrm{robust}, j}(t) = \sum_{k=1}^K \widehat{\omega}_{\text{stable},k} \, \widehat{X}_j^{(k)}(t), \quad j \in [p]. 
\end{align}

After obtaining $\widehat{X}_{\mathrm{robust}, j}(t)$, we next apply the kernel ODE estimation method \citep{dai2022kernel} to obtain the robust estimator of the link function $\widehat{F}(\cdot,t)$. Actually, this final step is flexible, and can adopt a variety of modeling approaches, for instance, parametric linear ODEs \citep{lu2011high, zhang2015dynamic}, nonparametric kernel ODEs \citep{dai2022kernel}, and highly flexible neural ODEs \citep{chen2018neural}.

To further illustrate the stability issue, Figure \ref{diff_dn_dist} reports the loss \(\|\widehat{\omega} - \omega^*\|\) of the naive plug-in estimator and the proposed estimator with different values of \(d_n\) under a simple example. In this example, we set \(\Gamma = \text{diag}\!\left( \begin{pmatrix} 1 & 1 \\ 1 & 1 \end{pmatrix},\,0,\,\begin{pmatrix} 2 & -2 \\ -2 & 2 \end{pmatrix} \right)\), and \(\widehat{\Gamma} = \Gamma + \big( z_{jj'}^2 \big)_{1 \le j,j' \le 5}\), where \(z_{jj'}\sim N(0,1/\sqrt{n})\) for \(j<j'\), and \(z_{jj'} = z_{j'j}\). We increase the number of time points \(n\) from 10 to 20,000. We consider three different values of \(d_n\): \(1 / n^2\), which decays rapidly and may be too fast to adequately account for estimation error in \(\widehat{\Gamma}\); \(\log(n) / n\), which decays moderately and aligns with theoretical suggestions for balancing error control; and \(1 / \log(n)\), which decays slowly and may lead to over-relaxation of the constraint. As observed in the figure, the naive plug-in estimator and the proposed estimator with \(d_n = 1/n^2\) exhibit persistent oscillation, with the loss failing to converge to zero even as \(n\) increases to infinity, due to insufficient tolerance for near-singularity and estimation error. In contrast, \(d_n = \log(n)/n\) leads to a smooth reduction in loss that approaches zero, demonstrating effective stabilization. Meanwhile, \(d_n = 1/\log(n)\) results in a smooth decrease but plateaus without reaching zero, indicating over-relaxation. These patterns suggest that \(d_n\) should be chosen with a proper rate to match the scale of estimation error and ensure consistent convergence, as guided by our theoretical analysis. We discuss in more detail the choice of the tolerance value $d_n$ in Sections \ref{sec:theory} and \ref{sec:simulations}.

\subsection{Inference for trajectory reconstruction}

We next study the inference, i.e., to construct a pointwise confidence interval for each component of the robust trajectory at any time $t$. 

For each individual trajectory estimator $\widehat{X}_j^{(k)}(t)$, we construct its confidence interval as,
\begin{align*} 
\left[ \widehat{X}_j^{(k)}(t) - z_{1-\alpha/2} \frac{\widehat{\sigma}_j^{(k)}(t)}{\sqrt{n h}}, \;\; 
\widehat{X}_j^{(k)}(t) + z_{1-\alpha/2} \frac{\widehat{\sigma}_j^{(k)}(t)}{\sqrt{n h}} \right],
\end{align*}
where $h$ is the bandwidth of the local polynomial regression or kernel smoother, $z_{1-\alpha/2}$ is the upper $\alpha/2$ quantile of the standard normal distribution, and $\widehat{\sigma}_j^{(k)}(t)$ is an estimate of the standard error for $\widehat{X}_j^{(k)}(t)$ at time $t$. This standard error is in turn estimated by
\begin{align*}
\widehat{\sigma}_j^{(k)}(t) = \left\{ \frac{1}{n h_{\mathrm{se}}} \sum_{i=1}^n G\left( \frac{t_i - t}{h_{\mathrm{se}}} \right)^2 \left(Y_{j}^{(k)}(t_i) - \widehat{X}_j^{(k)}(t_i) \right)^2 \right\}^{1/2},
\end{align*}
where $G(\cdot)$ is a kernel function, e.g., the Epanechnikov kernel, or the Gaussian kernel, and $h_{\mathrm{se}}$ is the bandwidth used for standard error estimation. 

It is noteworthy that the two bandwidths, $h$ and $h_{\mathrm{se}}$, play different roles. The bandwidth $h$ is used for estimating the trajectory. For a valid inference, $h$ should satisfy that $h \to 0$, $n h \to \infty$, and $n h^{2s+1} \to 0$, as $n \to \infty$, where $s$ is the H\"{o}lder smoothness order of the underlying trajectory. This ensures that the bias is negligible compared to the standard error, i.e., $E|\widehat{X}_j^{(k)}(t) - X_j^{(k)}(t)| = o_p(1/\sqrt{nh})$, and consequently $\sqrt{nh}(\widehat{X}_j^{(k)}(t) - X_j^{(k)}(t))$ is asymptotically normal with mean zero. The bandwidth $h_{\mathrm{se}}$ is used for estimating the standard error. It should also shrink to zero, typically at a rate such that $n h_{\mathrm{se}} \to \infty$ and $n h_{\mathrm{se}}^{2s+1} \to 0$, to ensure the standard error estimator itself is sufficiently accurate. In our setting, we require $|\widehat{\sigma}_j^{(k)}(t) - \sigma_j^{(k)}(t)| = o_p(1/\sqrt{n h})$, so that the error in standard error estimation does not affect the first-order validity of the confidence interval. These rate requirements are standard in nonparametric inference and are justified formally in Section \ref{sec:theory}.

Next, for the robust estimator $\widehat{X}_{\text{robust}, j}(t)$ in \eqref{eqn:robust-est}, its standard error is estimated by 
\begin{align*}
\widehat{\sigma}_{\text{robust}, j}(t) = \left\{ \sum_{k=1}^K \widehat{\omega}_{\text{stable},k}^2 \left(\widehat{\sigma}_j^{(k)}(t)\right)^2 \right\}^{1/2},
\end{align*}
where the cross-covariances can be ignored since different subjects or data sources are typically independent. This estimator remains valid as long as the stabilized weights $\widehat{\omega}_{\text{stable}}$ are estimated sufficiently accurately, specifically at a rate $\|\widehat{\omega}_{\text{stable}} - \omega_{\text{stable}}^*\| = o_p(1/\sqrt{n h})$.

The final confidence interval for the robust trajectory is then 
\begin{align}\label{combined-CI}
\left[
\widehat{X}_{\text{robust}, j}(t) - z_{1-\alpha/2} \frac{\widehat{\sigma}_{\text{robust}, j}(t)}{\sqrt{n h}}, \;\; 
\widehat{X}_{\text{robust}, j}(t) + z_{1-\alpha/2} \frac{\widehat{\sigma}_{\text{robust}, j}(t)}{\sqrt{n h}}
\right].
\end{align}
The validity of this confidence interval follows from the fact that, under suitable conditions, the bias of each nonparametric estimator and the estimation error of the standard error are both asymptotically negligible, and the weights are estimated consistently at a sufficiently fast rate. Therefore, the robust estimator is asymptotically normal and the plug-in standard error formula together yield a confidence interval with asymptotically correct coverage. These arguments and the detailed rate conditions are formalized in Section \ref{sec:theory}.

\section{Theory}
\label{sec:theory}

We establish theoretical guarantees for our proposed method, including the convergence property of the estimated $\widehat{\Gamma}$ and $\widehat{\omega}_{\text{stable}}$, the error bound for the robust estimator of the trajectory in \eqref{eqn:robust-est}, and the validity of the confidence interval in \eqref{combined-CI}. 

We begin with the convergence property of $\widehat{\Gamma}$ and $\widehat{\omega}_{\text{stable}}$. 

\begin{Proposition}\label{F-norm-Gamma-Err}
Suppose that each $X_j^{(k)}(t)$ is order-$s$ H\"{o}lder smooth for some $s > 1$, and the bandwidth $h$ satisfies that $h \asymp n^{-1/(2s+1)}$. Then,
\begin{align*}
\|\widehat{\Gamma} - \Gamma\|_F = O_p\left(p\, n^{-\frac{s-1}{2 s+1}}\right).
\end{align*}
\end{Proposition}

Proposition \ref{F-norm-Gamma-Err} obtains the convergence rate of $\widehat{\Gamma}$, which is standard and optimal in nonparametric regressions, depending only on the smoothness parameter $s$ and the dimension $p$. Notably, this result only requires a mild smoothness condition on the underlying signal trajectories, and places no additional structural assumption. In contrast, existing literature, such as \citet{guo2023statistical,wang2023distributionally}, often assumes non-degeneracy or additional structure on $\Gamma$ to ensure stability. 

\begin{Proposition}\label{thm2}
Suppose the same conditions as in Proposition \ref{F-norm-Gamma-Err} hold. Furthermore, suppose that the tolerance value $d_n$ satisfies that 
\begin{align}\label{dn-range}
2 \|\widehat{\Gamma} - \Gamma\|_{op}  \leq d_n. 
\end{align}
Let $\lambda_{\min}^+(\Gamma)$ denote the smallest positive eigenvalue of $\Gamma$. Then,
\begin{align*}
\left\|\widehat{\omega}_{\text{stable}} - \omega_{\text{stable}}^*\right\| 
\leq \sqrt{2^{\frac{3}{2}} \left(\frac{d_n}{\lambda_{\min}^+(\Gamma)}\right)^{1/2} + \frac{2 d_n}{\lambda_{\min}^+(\Gamma)}}.
\end{align*}  
\end{Proposition}

Proposition \ref{thm2} shows that the estimation error of the stabilized weight is controlled directly by $d_n$ and the smallest positive eigenvalue of \(\Gamma\). As long as $d_n$ dominates twice the spectral error in $\widehat{\Gamma}$, the estimator $\widehat{\omega}_{\text{stable}}$ is consistent, regardless of whether $\Gamma$ is positive definite or semi-definite. This reflects a strength of our method, i.e., it remains robust even when $\Gamma$ is or nearly singular. Additionally, while Proposition \ref{thm2} provides a non-asymptotic bound, it can be simplified in the asymptotic sense. In particular, if \eqref{dn-range} holds for a sufficiently large $n$ and $d_n/\lambda_{\min}^+(\Gamma)=o(1)$, then, 
\begin{align}\label{prop2-equiv-bound}
\|\widehat{\omega}_{\text{stable}} - \omega_{\text{stable}}^*\| \lesssim \left( \frac{d_n}{\lambda_{\min}^+(\Gamma)} \right)^{1/4}. 
\end{align}
This convergence rate is slower than the parametric $n^{-1 / 2}$ rate, reflecting the nonparametric nature of our problem, because both the latent trajectories $\left\{X^{(k)}(t)\right\}$ and their derivatives $\left\{D_t X^{(k)}(t)\right\}$ must be estimated nonparametrically from noisy observations before the construction of the estimators for $\Gamma$ and $\omega$. 

We further discuss the choice of the tolerance value $d_n$. By Proposition \ref{F-norm-Gamma-Err}, when we choose $d_n \succ p\, n^{-\frac{s-1}{2s+1}}$, the condition in \eqref{dn-range} holds with a high probability. In practice, however, we do not know the smoothness parameter $s$. We thus propose a sample-splitting-based approach to adaptively choose $d_n$. More specifically, we partition the training data into two parts $\mathcal I_1$ and $\mathcal I_2$ with approximately equal size. We estimate $\widehat{\Gamma}_{\mathcal I_1}$ and $\widehat{\Gamma}_{\mathcal I_2}$ using each half, and set 
\begin{align} \label{eqn:dn}
d_n = C_{d}\,\log(n)\,
\left( \frac{\big\|\widehat{\Gamma}_{\mathcal I_1}-\widehat{\Gamma}_{\mathcal I_2}\big\|_{op}}{\|\widehat{\Gamma}\|_{op}} \land 1 \right)
\end{align}
where $C_{d}$ is a constant and is recommended to take the value between $0.01$ and $0.1$, $\log(n)$ is for asymptotic dominance, $\big\|\widehat{\Gamma}_{\mathcal I_1}-\widehat{\Gamma}_{\mathcal I_2}\big\|_{op}$ in the nominator is an approximation of $\|\widehat{\Gamma} - \Gamma\|_{op}$, $\|\widehat{\Gamma}\|_{op}$ in the denominator is for scaling, and taking minimization with 1 is to truncate potential outliers. We study the empirical performance of this choice in Section \ref{sec:simulations}. 

Next, we establish the estimation error of our robust trajectory estimator and the validity of the corresponding confidence interval. 

\begin{Theorem} \label{est-rob}
Suppose the same conditions as in Proposition \ref{F-norm-Gamma-Err} hold. Then, 
\begin{align*}
\int_0^T \left( \sum_{k=1}^K \widehat{\omega}_{k} \widehat{X}_j^{(k)}(t) - \sum_{k=1}^K \omega_{\text{stable},k}^* X_j^{(k)}(t) \right)^2 dt 
= O_p\left(K^2 n^{-\frac{s}{2 s+1}} \vee K\|\widehat{\omega}^*- \omega_{\text{stable}}^*\|^2\right).
\end{align*}
\end{Theorem}

Theorem \ref{est-rob} shows that the estimation error is controlled by two sources: the nonparametric trajectory estimation error $K^2 n^{-\frac{s}{2 s+1}}$ that arises from estimating $\Gamma$, and the weight estimation error $K\|\widehat{\omega}^*- \omega_{\text{stable}}^*\|^2$ that arises from estimating the stabilized weights. When the weights are estimated consistently, the estimation error is bounded by the nonparametric rate for the trajectory estimation, reflecting how our method aggregates information across $K$ sources while maintaining efficiency under data heterogeneity.

\begin{Theorem}\label{thm-inference}
Suppose that each $X_j^{(k)}(t)$ is order-$s$ H\"older smooth for some $s>1$. Suppose $d_n>0$ and $\|\widehat{\Gamma}-\Gamma\|_{op}\prec d_n = o(1)$. Suppose the bandwidth $h$ satisfies that 
\begin{align} \label{eq:h-cond-inference}
n^{-1} \prec h \prec n^{-\frac{1}{2s+1}},
\qquad
(nh)^2 d_n \to 0.
\end{align}
Then the confidence interval in \eqref{combined-CI} achieves the nominal coverage probability asymptotically; that is, for each fixed \(t\in[0,T]\) and \(j\in[p]\),
\begin{align*}
P\!\left(\sum_{k=1}^K \omega_{stable,k}^* X_j^{(k)}(t) \in \mathrm{CI}_j(t) \right) \to 1-\alpha,
\end{align*}
where \(\mathrm{CI}_j(t)\) denotes the interval in \eqref{combined-CI}.
\end{Theorem}

Theorem \ref{thm-inference} shows that our confidence interval in \eqref{combined-CI} is asymptotically valid. Its length is 
\begin{align*}
\text{Length}\big(\mathrm{CI}_j(t)\big) = 2 z_{1-\alpha/2}\,
\frac{
\sqrt{ \sum_{k=1}^K \big(\widehat{\omega}_{stable,k}\big)^2
\big(\widehat{\sigma}_j^{(k)}(t)\big)^2 }
}{\sqrt{nh}}.
\end{align*}
Under the conditions of Theorem~\ref{thm-inference}, there exists a constant $C>0$, such that
\begin{align*}
\lim_{n\to\infty}
P\!\left(
\text{Length}\big(\mathrm{CI}_j(t)\big)\le \frac{C}{\sqrt{nh}}
\right)=1.
\end{align*}
Therefore the interval length is of order $(nh)^{-1/2}$. Moreover, the largest admissible bandwidth is of order $n^{-\frac{1}{2s+1}} \wedge \big(n^{-1} d_n^{-1/2}\big)$, so the shortest achievable interval length, up to some constant, is of order
\begin{align*}
\mathrm{Length}\big(\mathrm{CI}_j(t)\big) \asymp n^{-\frac{s}{2s+1}} \,\vee\, d_n^{1/4}.
\end{align*}
The additional term $d_n^{1/4}$ here reflects the extra undersmoothing required to make the stabilized weight estimation error asymptotically negligible.

We further discuss the bandwidth conditions in \eqref{eq:h-cond-inference}. The condition that $n^{-1}\prec h$ implies $nh\to\infty$, which is necessary for the pointwise asymptotic normality of the local polynomial estimator. The condition that $h\prec n^{-1/(2s+1)}$ implies $n h^{2s+1}\to 0$, which makes the smoothing bias negligible compared to the stochastic fluctuation of order $(nh)^{-1/2}$. The condition $(nh)^2 d_n\to 0$ ensures that the bias from estimating the stabilized weights is asymptotically negligible. In particular, the result in  \eqref{prop2-equiv-bound} and $(nh)^2 d_n\to 0$ together imply that $\|\widehat{\omega}_{stable}-\omega_{stable}^*\|_2 = o_p\!\big((nh)^{-1/2}\big)$, suggesting that the weight estimation error is negligible compared to the standard deviation $(nh)^{-1/2}$, thus not affecting the asymptotic distribution of the robust trajectory estimator.

\section{Simulation Studies}
\label{sec:simulations}

\subsection{Simulation setup}

We consider two common ODE systems to evaluate the empirical performance of our proposed method and compare it with some alternative solutions. 

\begin{Example}\label{eg-NFBLB}
The first example is a three-node enzyme regulatory network of a negative feedback loop with a buffering node \citep{ma2009defining}, defined as, 
\begin{align*} 
\begin{aligned}
\frac{d X_1(t)}{d t} & =c_1 \frac{c_0\left\{1-X_1(t)\right\}}{\left\{1-X_1(t)\right\}+C_1}-\tilde{c}_1 c_2 \frac{X_1(t)}{X_1(t)+C_2}, \\
\frac{d X_2(t)}{d t} & =c_3 \frac{\left\{1-X_2(t)\right\} X_3(t)}{\left\{1-X_2(t)\right\}+C_3}-\tilde{c}_2 c_4 \frac{X_2(t)}{X_2(t)+C_4}, \\
\frac{d X_3(t)}{d t} & =c_5 \frac{X_1(t)\left\{1-X_3(t)\right\}}{\left\{1-X_3(t)\right\}+C_5}-c_6 \frac{X_2(t) X_3(t)}{X_3(t)+C_6},
\end{aligned}
\end{align*}
where $X_1(t), X_2(t), X_3(t)$ are three interacting nodes, such that $X_1(t)$ receives the input, $X_2(t)$ plays the diverse regulatory role, and $X_3(t)$ transmits the output, $c_0$ is the initial input stimulus, and $c_1, \ldots, c_6, C_1, \ldots, C_6, \tilde{c}_1, \tilde{c}_2$ denote the catalytic rate parameters, the Michaelis-Menten constants, and the concentration parameters, respectively.
\end{Example}

\begin{Example}\label{eg-LV}
The second example is the ten-dimensional Lotka-Volterra equations, which are pairs of first-order nonlinear differential equations describing the dynamics of biological systems in
which predators and prey interact \citep{volterra1928variations}, defined as, 
\begin{align*} 
\begin{aligned}
\frac{d X_{2 j-1}(t)}{d t} & =\alpha_{1, j} X_{2 j-1}(t)-\alpha_{2, j} X_{2 j-1}(t) X_{2 j}(t), \quad j=1,\ldots,5, \\
\frac{d X_{2 j}(t)}{d t} & = \alpha_{3, j} X_{2 j-1}(t) X_{2 j}(t)-\alpha_{4, j} X_{2 j}(t),
\end{aligned}
\end{align*}
where the parameters $\alpha_{2, j}$ and $\alpha_{3, j}$ define the interaction between the two populations such that $d X_{2 j-1}(t) / d t$ and $d X_{2 j}(t) / d t$ are non-additive functions of $X_{2 j-1}$ and $X_{2 j}$, with $X_{2 j-1}$ denoting the prey and $X_{2 j}$ the predator.
\end{Example}

Based on the above ODE systems, we next introduce different levels of subject-specific heterogeneity. We consider $K=\{5,10\}$ subjects, and observe the data from,
\begin{align*}
Y^{(k)}(t)=X^{(k)}(t)+\epsilon^{(k)}(t), \quad t=t_1, \ldots, t_n, \; 0 \leq t_1< \ldots t_n \leq T, \; k=1,\ldots,K. 
\end{align*}
Following \citet{dai2022kernel}, we set $n=40, p=3, T=2, \text{var}(\epsilon_j^{(k)}(t)) = 0.01^2$ for Example \ref{eg-NFBLB}, and set $n=200, p=10, T=100, \text{var}(\epsilon_j^{(k)}(t)) = 1$ for Example \ref{eg-LV}. Moreover, for Example \ref{eg-NFBLB}, we set the initial state $X_j^{(k)}(0)=0.5$, and set the ODE parameters as, 
\begin{enumerate}[Level I: ]
\item $c_0 = 1, c_{1}=c_{2}=c_{3}=c_{5}=c_{6}=10, c_{4}=1,C_{1}=\ldots=C_{6}=0.1,\tilde{c}_{1}=1, \tilde{c}_{2}=0.2$;
\item $c_0 = 1+k/40, c_{1}=c_{2}=c_{3}=c_{5}=c_{6}=10 (1+k/40), c_{4}=1+k/40, C_{1}=\ldots=C_{6}=0.1 (1+k/40),\tilde{c}_{1}=1+k/40, \tilde{c}_{2}=0.2(1+k/40)$;
\item $c_0 = 1+k/20, c_{1}=c_{2}=c_{3}=c_{5}=c_{6}=10 (1+k/20), c_{4}=1+k/20, C_{1}=\ldots=C_{6}=0.1 (1+k/20),\tilde{c}_{1}=1+k/20, \tilde{c}_{2}=0.2(1+k/5)$;
\end{enumerate}
reflecting increasing level of heterogeneity. Similarly, for Example \ref{eg-LV}, we set the initial state $X_j^{(k)}(0)=1$, and set the ODE parameters as, 
\begin{enumerate}[Level I: ]
\item $\alpha_{1, j}=1.1+0.2(j-1), \alpha_{2, j}=0.4+0.2(j-1), \alpha_{3, j}=0.1+0.2(j-1), \alpha_{4, j}= 0.4+0.2(j-1)$;
\item $\alpha_{1, j}=\{1.1+0.2(j-1)\}(1+k/160), \alpha_{2, j}=\{0.4+0.2(j-1)\}(1+k/160), \alpha_{3, j}=\{0.1+0.2(j-1)\}(1+k/160), \alpha_{4, j}= \{0.4+0.2(j-1)\}(1+k/160)$;
\item $\alpha_{1, j}=\{1.1+0.2(j-1)\}(1+k/80), \alpha_{2, j}=\{0.4+0.2(j-1)\}(1+k/80), \alpha_{3, j}=\{0.1+0.2(j-1)\}(1+k/80), \alpha_{4, j}=\{0.4+0.2(j-1)\}(1+k/80)$; 
\end{enumerate}
for $j=1, \ldots, 5$, again reflecting increasing level of heterogeneity.

Moreover, we consider a stable case where we generate $K$ subject-specific systems following the above model settings, and an unstable case where we generate $(K-1)$ subject-specific systems following the above model settings, then generate the $K$th system by taking a linear combination of the first generated $(K-1)$ systems, and the weights are randomly sampled from a uniform distribution on a $(K-1)$-dimensional simplex.

\subsection{Estimation}\label{sec:experiments}

We first examine the estimation performance of our proposed method, and compare it with the empirical risk minimization method (ERM), and the invariance score method (IVS). The implementations of the latter two methods are based on \citet{dai2022kernel}. We choose $d_n$ following the data-adaptive way as outlined in \eqref{eqn:dn} where we set $C_d = 0.01$. We have experimented with other values of $C_d$ and obtained qualitatively similar results. We evaluate the estimation accuracy through various loss criteria, defined as,
\begin{align*}
\text{Max Loss: } & \max_{k\in[K]}\int_{0}^T \sum_{j=1}^p \left( D_t X_j^{(k)}(t)-\widetilde{F}_j(X^{(k)}(t),t) \right)^2 dt;\\
\text{Average Loss: } & \frac{1}{K}\sum_{k=1}^K\int_{0}^T \sum_{j=1}^p \left( D_t X_j^{(k)}(t)-\widetilde{F}_j(X^{(k)}(t),t) \right)^2 dt;\\
\text{Generalization Loss: } & \int_{0}^T \sum_{j=1}^p \left( D_t X_j^{(K+1)}(t)-\widetilde{F}_j(X^{(K+1)}(t),t) \right)^2 dt;
\end{align*}
where $\widehat{F}(\cdot,t)$ denotes the estimated link function based on the estimated signal trajectories $\widehat{X}_{\mathrm{robust}, j}(t)$. The first two loss functions measure the maximum discrepancy and the average discrepancy, respectively, between the estimated link function and the $K$ link functions in the training data, whereas the last loss function measures the discrepancy between the estimated link function and the $(K+1)$th link function in the future data. Tables \ref{tab1} and \ref{tab2} report the results for Examples \ref{eg-NFBLB} and \ref{eg-LV}, respectively, based on 100 data replications. 

\begin{table}[t!]
\centering
\caption{Estimation performance for Example \ref{eg-NFBLB}. Reported are the average max loss, average loss, and generalization loss, with the standard deviation in the parenthesis, based on 100 data replications. Considered models include the stable and unstable case, each with three levels of heterogeneity, and two values of number of subjects $K = \{5, 10\}$. Three solutions are compared, the proposed method, the empirical risk minimization method (ERM), and the invariance score method (IVS).}
\resizebox{\textwidth}{!}{%
\begin{tabular}{|ccc|ccc|ccc|}
\hline
\multicolumn{3}{|c|}{} 
  & \multicolumn{3}{c|}{$K=5$} 
  & \multicolumn{3}{c|}{$K=10$} \\ \hline
\multicolumn{1}{|c|}{Case} 
  & \multicolumn{1}{c|}{Level} 
  & Loss    
  & \multicolumn{1}{c|}{Proposed} 
  & \multicolumn{1}{c|}{ERM} 
  & IVS 
  & \multicolumn{1}{c|}{Proposed} 
  & \multicolumn{1}{c|}{ERM} 
  & IVS \\ \hline
\multicolumn{1}{|c|}{\multirow{9}{*}{Stable}} 
  & \multicolumn{1}{c|}{\multirow{3}{*}{I}} 
  & Max. 
  & \multicolumn{1}{c|}{0.068 (0.002)} & \multicolumn{1}{c|}{1.067 (0.002)} & 0.424 (0.003) 
  & \multicolumn{1}{c|}{0.066 (0.002)} & \multicolumn{1}{c|}{1.072 (0.001)} & 0.418 (0.004) \\ \cline{3-9}
\multicolumn{1}{|c|}{} & \multicolumn{1}{c|}{} 
  & Avg. 
  & \multicolumn{1}{c|}{0.068 (0.002)} & \multicolumn{1}{c|}{1.053 (0.002)} & 0.411 (0.003) 
  & \multicolumn{1}{c|}{0.066 (0.002)} & \multicolumn{1}{c|}{1.055 (0.001)} & 0.399 (0.004) \\ \cline{3-9}
\multicolumn{1}{|c|}{} & \multicolumn{1}{c|}{} 
  & Gen. 
  & \multicolumn{1}{c|}{0.068 (0.002)} & \multicolumn{1}{c|}{1.049 (0.001)} & 0.406 (0.003) 
  & \multicolumn{1}{c|}{0.066 (0.002)} & \multicolumn{1}{c|}{1.050 (0.001)} & 0.393 (0.004) \\ \cline{2-9}
\multicolumn{1}{|c|}{} 
  & \multicolumn{1}{c|}{\multirow{3}{*}{II}} 
  & Max. 
  & \multicolumn{1}{c|}{0.089 (0.003)} & \multicolumn{1}{c|}{1.246 (0.002)} & 0.454 (0.002) 
  & \multicolumn{1}{c|}{0.134 (0.003)} & \multicolumn{1}{c|}{1.433 (0.002)} & 0.527 (0.009) \\ \cline{3-9}
\multicolumn{1}{|c|}{} & \multicolumn{1}{c|}{} 
  & Avg. 
  & \multicolumn{1}{c|}{0.076 (0.002)} & \multicolumn{1}{c|}{1.197 (0.002)} & 0.439 (0.002) 
  & \multicolumn{1}{c|}{0.095 (0.002)} & \multicolumn{1}{c|}{1.328 (0.001)} & 0.500 (0.008) \\ \cline{3-9}
\multicolumn{1}{|c|}{} & \multicolumn{1}{c|}{} 
  & Gen.
  & \multicolumn{1}{c|}{0.097 (0.003)} & \multicolumn{1}{c|}{1.264 (0.001)} & 0.445 (0.002) 
  & \multicolumn{1}{c|}{0.145 (0.003)} & \multicolumn{1}{c|}{1.448 (0.001)} & 0.517 (0.008) \\ \cline{2-9}
\multicolumn{1}{|c|}{} 
  & \multicolumn{1}{c|}{\multirow{3}{*}{III}} 
  & Max. 
  & \multicolumn{1}{c|}{0.122 (0.003)} & \multicolumn{1}{c|}{1.442 (0.002)} & 0.494 (0.003) 
  & \multicolumn{1}{c|}{0.233 (0.004)} & \multicolumn{1}{c|}{1.793 (0.002)} & 0.850 (0.007) \\ \cline{3-9}
\multicolumn{1}{|c|}{} & \multicolumn{1}{c|}{} 
  & Avg. 
  & \multicolumn{1}{c|}{0.093 (0.003)} & \multicolumn{1}{c|}{1.348 (0.002)} & 0.470 (0.002) 
  & \multicolumn{1}{c|}{0.137 (0.003)} & \multicolumn{1}{c|}{1.609 (0.002)} & 0.763 (0.007) \\ \cline{3-9}
\multicolumn{1}{|c|}{} & \multicolumn{1}{c|}{} 
  & Gen. 
  & \multicolumn{1}{c|}{0.143 (0.003)} & \multicolumn{1}{c|}{1.479 (0.002)} & 0.505 (0.003) 
  & \multicolumn{1}{c|}{0.260 (0.005)} & \multicolumn{1}{c|}{1.823 (0.002)} & 0.831 (0.006) \\ \hline
\multicolumn{1}{|c|}{\multirow{9}{*}{Unstable}} 
  & \multicolumn{1}{c|}{\multirow{3}{*}{I}} 
  & Max.  
  & \multicolumn{1}{c|}{0.067 (0.002)} & \multicolumn{1}{c|}{1.065 (0.002)} & 0.421 (0.004) 
  & \multicolumn{1}{c|}{0.068 (0.002)} & \multicolumn{1}{c|}{1.072 (0.001)} & 0.420 (0.004) \\ \cline{3-9}
\multicolumn{1}{|c|}{} & \multicolumn{1}{c|}{} 
  & Avg. 
  & \multicolumn{1}{c|}{0.067 (0.002)} & \multicolumn{1}{c|}{1.051 (0.001)} & 0.408 (0.003) 
  & \multicolumn{1}{c|}{0.068 (0.002)} & \multicolumn{1}{c|}{1.053 (0.001)} & 0.401 (0.004) \\ \cline{3-9}
\multicolumn{1}{|c|}{} & \multicolumn{1}{c|}{} 
  & Gen.  
  & \multicolumn{1}{c|}{0.067 (0.002)} & \multicolumn{1}{c|}{1.046 (0.001)} & 0.403 (0.003) 
  & \multicolumn{1}{c|}{0.068 (0.002)} & \multicolumn{1}{c|}{1.048 (0.001)} & 0.395 (0.004) \\ \cline{2-9}
\multicolumn{1}{|c|}{} 
  & \multicolumn{1}{c|}{\multirow{3}{*}{II}} 
  & Max. 
  & \multicolumn{1}{c|}{0.087 (0.002)} & \multicolumn{1}{c|}{1.209 (0.002)} & 0.441 (0.003) 
  & \multicolumn{1}{c|}{0.125 (0.003)} & \multicolumn{1}{c|}{1.393 (0.002)} & 0.498 (0.006) \\ \cline{3-9}
\multicolumn{1}{|c|}{} & \multicolumn{1}{c|}{} 
  & Avg.  
  & \multicolumn{1}{c|}{0.078 (0.002)} & \multicolumn{1}{c|}{1.171 (0.002)} & 0.426 (0.003) 
  & \multicolumn{1}{c|}{0.092 (0.002)} & \multicolumn{1}{c|}{1.298 (0.001)} & 0.474 (0.005) \\ \cline{3-9}
\multicolumn{1}{|c|}{} & \multicolumn{1}{c|}{} 
  & Gen.  
  & \multicolumn{1}{c|}{0.103 (0.003)} & \multicolumn{1}{c|}{1.251 (0.001)} & 0.436 (0.003) 
  & \multicolumn{1}{c|}{0.147 (0.003)} & \multicolumn{1}{c|}{1.430 (0.001)} & 0.497 (0.005) \\ \cline{2-9}
\multicolumn{1}{|c|}{} 
  & \multicolumn{1}{c|}{\multirow{3}{*}{III}} 
  & Max.  
  & \multicolumn{1}{c|}{0.109 (0.003)} & \multicolumn{1}{c|}{1.361 (0.002)} & 0.466 (0.003) 
  & \multicolumn{1}{c|}{0.196 (0.004)} & \multicolumn{1}{c|}{1.718 (0.002)} & 0.829 (0.008) \\ \cline{3-9}
\multicolumn{1}{|c|}{} & \multicolumn{1}{c|}{} 
  & Avg. 
  & \multicolumn{1}{c|}{0.088 (0.002)} & \multicolumn{1}{c|}{1.289 (0.002)} & 0.449 (0.003) 
  & \multicolumn{1}{c|}{0.117 (0.003)} & \multicolumn{1}{c|}{1.547 (0.002)} & 0.751 (0.007) \\ \cline{3-9}
\multicolumn{1}{|c|}{} & \multicolumn{1}{c|}{} 
  & Gen.  
  & \multicolumn{1}{c|}{0.150 (0.003)} & \multicolumn{1}{c|}{1.445 (0.002)} & 0.497 (0.002) 
  & \multicolumn{1}{c|}{0.248 (0.005)} & \multicolumn{1}{c|}{1.786 (0.002)} & 0.832 (0.006) \\ \hline
\end{tabular}}
\label{tab1}
\end{table}

We make some observations. First, our proposed method consistently outperforms the competing approaches across all settings. The improvements are substantial and are more pronounced in Example \ref{eg-LV}, which involves a $10$-dimensional dynamic system, than in Example \ref{eg-NFBLB}, which involves only a $3$-dimensional system. For instance, in Example \ref{eg-LV}, the proposed method achieves max loss values that are 25 to 115 times smaller than those produced by ERM and IVS. Similar trends are observed for the average loss and generalization loss, where our method attains losses that are one to two orders of magnitude smaller than the competing methods. Second, as the heterogeneity level increases from I to III, or as the number of subjects $K$ increases from 5 to 10, the problem becomes more challenging and the performance of all methods tends to deteriorate. Nevertheless, our method exhibits the smallest degradation in performance compared to ERM and IVS. For instance, in Example \ref{eg-NFBLB}, the increase in max loss for the proposed method is substantially smaller, by several-fold, than the increases observed for ERM and IVS. Finally, our method attains a similar performance for both the stable and unstable cases, confirming that our stabilized estimator is robust to the degeneracy of the $\Gamma$ matrix. 

\begin{table}[t!]
\centering
\caption{Estimation performance for Example \ref{eg-LV}. Reported are the average max loss, average loss, and generalization loss ($\times 10^4$), with the standard deviation in the parenthesis, based on 100 data replications. Considered models include the stable and unstable case, each with three levels of heterogeneity, and two values of number of subjects $K = \{5, 10\}$. Three solutions are compared, the proposed method, the empirical risk minimization method (ERM), and the invariance score method (IVS).}
\resizebox{\textwidth}{!}{%
\begin{tabular}{|ccc|ccc|ccc|}
\hline
\multicolumn{3}{|c|}{}
  & \multicolumn{3}{c|}{$K=5$}
  & \multicolumn{3}{c|}{$K=10$} \\ \hline
\multicolumn{1}{|c|}{Case}
  & \multicolumn{1}{c|}{Level}
  & Loss    
  & \multicolumn{1}{c|}{Proposed}
  & \multicolumn{1}{c|}{ERM}
  & IVS
  & \multicolumn{1}{c|}{Proposed}
  & \multicolumn{1}{c|}{ERM}
  & IVS \\ \hline
\multicolumn{1}{|c|}{\multirow{9}{*}{Stable}}
  & \multicolumn{1}{c|}{\multirow{3}{*}{I}}
  & Max.
  & \multicolumn{1}{c|}{0.044 (0.000)} & \multicolumn{1}{c|}{4.843 (0.001)} & 4.798 (0.006)
  & \multicolumn{1}{c|}{0.043 (0.000)} & \multicolumn{1}{c|}{4.851 (0.001)} & 4.775 (0.003) \\ \cline{3-9}
\multicolumn{1}{|c|}{} & \multicolumn{1}{c|}{}
  & Avg.  
  & \multicolumn{1}{c|}{0.044 (0.000)} & \multicolumn{1}{c|}{4.838 (0.001)} & 4.790 (0.006)
  & \multicolumn{1}{c|}{0.043 (0.000)} & \multicolumn{1}{c|}{4.845 (0.001)} & 4.766 (0.003) \\ \cline{3-9}
\multicolumn{1}{|c|}{} & \multicolumn{1}{c|}{}
  & Gen.  
  & \multicolumn{1}{c|}{0.044 (0.000)} & \multicolumn{1}{c|}{4.758 (0.002)} & 4.708 (0.008)
  & \multicolumn{1}{c|}{0.043 (0.000)} & \multicolumn{1}{c|}{4.770 (0.001)} & 4.675 (0.004) \\ \cline{2-9}
\multicolumn{1}{|c|}{}
  & \multicolumn{1}{c|}{\multirow{3}{*}{II}}
  & Max.
  & \multicolumn{1}{c|}{0.162 (0.000)} & \multicolumn{1}{c|}{6.108 (0.001)} & 6.038 (0.006)
  & \multicolumn{1}{c|}{0.238 (0.000)} & \multicolumn{1}{c|}{7.522 (0.001)} & 7.417 (0.004) \\ \cline{3-9}
\multicolumn{1}{|c|}{} & \multicolumn{1}{c|}{}
  & Avg.  
  & \multicolumn{1}{c|}{0.139 (0.000)} & \multicolumn{1}{c|}{5.258 (0.001)} & 5.191 (0.006)
  & \multicolumn{1}{c|}{0.209 (0.000)} & \multicolumn{1}{c|}{5.534 (0.001)} & 5.427 (0.003) \\ \cline{3-9}
\multicolumn{1}{|c|}{} & \multicolumn{1}{c|}{}
  & Gen.  
  & \multicolumn{1}{c|}{0.219 (0.001)} & \multicolumn{1}{c|}{4.787 (0.002)} & 4.719 (0.008)
  & \multicolumn{1}{c|}{0.287 (0.001)} & \multicolumn{1}{c|}{5.673 (0.001)} & 5.546 (0.005) \\ \cline{2-9}
\multicolumn{1}{|c|}{}
  & \multicolumn{1}{c|}{\multirow{3}{*}{III}}
  & Max.
  & \multicolumn{1}{c|}{0.230 (0.000)} & \multicolumn{1}{c|}{7.515 (0.001)} & 7.452 (0.006)
  & \multicolumn{1}{c|}{0.308 (0.000)} & \multicolumn{1}{c|}{9.146 (0.001)} & 9.029 (0.006) \\ \cline{3-9}
\multicolumn{1}{|c|}{} & \multicolumn{1}{c|}{}
  & Avg.  
  & \multicolumn{1}{c|}{0.205 (0.000)} & \multicolumn{1}{c|}{5.724 (0.001)} & 5.657 (0.006)
  & \multicolumn{1}{c|}{0.276 (0.000)} & \multicolumn{1}{c|}{6.461 (0.001)} & 6.374 (0.004) \\ \cline{3-9}
\multicolumn{1}{|c|}{} & \multicolumn{1}{c|}{}
  & Gen.
  & \multicolumn{1}{c|}{0.340 (0.001)} & \multicolumn{1}{c|}{6.880 (0.002)} & 6.803 (0.008)
  & \multicolumn{1}{c|}{0.363 (0.001)} & \multicolumn{1}{c|}{8.214 (0.002)} & 8.090 (0.007) \\ \hline
\multicolumn{1}{|c|}{\multirow{9}{*}{Unstable}}
  & \multicolumn{1}{c|}{\multirow{3}{*}{I}}
  & Max.
  & \multicolumn{1}{c|}{0.043 (0.000)} & \multicolumn{1}{c|}{4.841 (0.001)} & 4.857 (0.005)
  & \multicolumn{1}{c|}{0.042 (0.000)} & \multicolumn{1}{c|}{4.851 (0.001)} & 4.830 (0.004) \\ \cline{3-9}
\multicolumn{1}{|c|}{} & \multicolumn{1}{c|}{}
  & Avg.
  & \multicolumn{1}{c|}{0.043 (0.000)} & \multicolumn{1}{c|}{4.835 (0.001)} & 4.850 (0.005)
  & \multicolumn{1}{c|}{0.042 (0.000)} & \multicolumn{1}{c|}{4.845 (0.001)} & 4.821 (0.004) \\ \cline{3-9}
\multicolumn{1}{|c|}{} & \multicolumn{1}{c|}{}
  & Gen.  
  & \multicolumn{1}{c|}{0.043 (0.000)} & \multicolumn{1}{c|}{4.755 (0.002)} & 4.797 (0.007)
  & \multicolumn{1}{c|}{0.042 (0.000)} & \multicolumn{1}{c|}{4.769 (0.001)} & 4.759 (0.005) \\ \cline{2-9}
\multicolumn{1}{|c|}{}
  & \multicolumn{1}{c|}{\multirow{3}{*}{II}}
  & Max.
  & \multicolumn{1}{c|}{0.134 (0.000)} & \multicolumn{1}{c|}{5.312 (0.001)} & 5.310 (0.006)
  & \multicolumn{1}{c|}{0.226 (0.000)} & \multicolumn{1}{c|}{7.526 (0.001)} & 7.465 (0.004) \\ \cline{3-9}
\multicolumn{1}{|c|}{} & \multicolumn{1}{c|}{}
  & Avg.  
  & \multicolumn{1}{c|}{0.095 (0.000)} & \multicolumn{1}{c|}{4.432 (0.006)} & 4.430 (0.008)
  & \multicolumn{1}{c|}{0.184 (0.000)} & \multicolumn{1}{c|}{5.063 (0.002)} & 5.002 (0.004) \\ \cline{3-9}
\multicolumn{1}{|c|}{} & \multicolumn{1}{c|}{}
  & Gen.  
  & \multicolumn{1}{c|}{0.299 (0.001)} & \multicolumn{1}{c|}{4.784 (0.002)} & 4.806 (0.008)
  & \multicolumn{1}{c|}{0.339 (0.001)} & \multicolumn{1}{c|}{5.677 (0.001)} & 5.615 (0.006) \\ \cline{2-9}
\multicolumn{1}{|c|}{}
  & \multicolumn{1}{c|}{\multirow{3}{*}{III}}
  & Max.
  & \multicolumn{1}{c|}{0.200 (0.001)} & \multicolumn{1}{c|}{7.520 (0.002)} & 7.517 (0.007)
  & \multicolumn{1}{c|}{0.303 (0.000)} & \multicolumn{1}{c|}{7.519 (0.001)} & 7.467 (0.004) \\ \cline{3-9}
\multicolumn{1}{|c|}{} & \multicolumn{1}{c|}{}
  & Avg.  
  & \multicolumn{1}{c|}{0.154 (0.000)} & \multicolumn{1}{c|}{4.938 (0.009)} & 4.929 (0.011)
  & \multicolumn{1}{c|}{0.247 (0.000)} & \multicolumn{1}{c|}{5.630 (0.002)} & 5.575 (0.004) \\ \cline{3-9}
\multicolumn{1}{|c|}{} & \multicolumn{1}{c|}{}
  & Gen.
  & \multicolumn{1}{c|}{0.439 (0.001)} & \multicolumn{1}{c|}{6.885 (0.002)} & 6.900 (0.009)
  & \multicolumn{1}{c|}{0.412 (0.001)} & \multicolumn{1}{c|}{8.234 (0.002)} & 8.160 (0.006) \\ \hline
\end{tabular}}
\label{tab2}
\end{table}

\subsection{Inference}

We next examine the inference performance of our proposed method. Since there is no existing alternative method, we focus on the empirical coverage probability (ECP) and confidence interval length (CIL) of our own method. For each data replication, we compute the coverage probabilities and the interval lengths at all sampled time points and across all dimensions, and record their averages, and we repeat with 100 data replications. We adopt similar example settings as in estimation. Table \ref{tab3} reports the results. 

\begin{table}[t!]
\centering
\caption{Inference performance for Examples \ref{eg-NFBLB} and \ref{eg-LV}. Reported are the empirical coverage probability (ECP) and confidence interval length (CIL) based on 100 data replications. Considered models include the stable and unstable case, each with three levels of heterogeneity, and two values of number of subjects $K = \{5, 10\}$.}
\begin{tabular}{|c|c|cc|cc|cc|cc|}
\hline
\multicolumn{2}{|c|}{} & \multicolumn{4}{c|}{Stable} & \multicolumn{4}{c|}{Unstable} \\ \hline
\multicolumn{2}{|c|}{} & \multicolumn{2}{c|}{$K=5$} & \multicolumn{2}{c|}{$K=10$} & \multicolumn{2}{c|}{$K=5$} & \multicolumn{2}{c|}{$K=10$} \\ \hline
\multicolumn{1}{|c|}{Example } & Level & \multicolumn{1}{c|}{ECP} & CIL & \multicolumn{1}{c|}{ECP} & CIL & \multicolumn{1}{c|}{ECP} & CIL & \multicolumn{1}{c|}{ECP} & CIL \\ \hline
\multicolumn{1}{|c|}{\multirow{3}{*}{Example \ref{eg-NFBLB}}} & I & \multicolumn{1}{c|}{97.0\%} & 0.469 & \multicolumn{1}{c|}{95.8\%} & 0.453 & \multicolumn{1}{c|}{96.2\%} & 0.476 & \multicolumn{1}{c|}{96.1\%} & 0.431 \\ \cline{2-10}
\multicolumn{1}{|c|}{} & II & \multicolumn{1}{c|}{96.9\%} & 0.499 & \multicolumn{1}{c|}{95.3\%} & 0.439 & \multicolumn{1}{c|}{96.7\%} & 0.472 & \multicolumn{1}{c|}{95.5\%} & 0.444 \\ \cline{2-10}
\multicolumn{1}{|c|}{} & III & \multicolumn{1}{c|}{96.7\%} & 0.453 & \multicolumn{1}{c|}{96.1\%} & 0.439 & \multicolumn{1}{c|}{96.4\%} & 0.450 & \multicolumn{1}{c|}{95.8\%} & 0.459 \\ \hline
\multicolumn{1}{|c|}{\multirow{3}{*}{Example \ref{eg-LV}}} & I & \multicolumn{1}{c|}{94.5\%} & 2.786 & \multicolumn{1}{c|}{94.9\%} & 2.490 & \multicolumn{1}{c|}{94.8\%} & 2.872 & \multicolumn{1}{c|}{94.9\%} & 2.582 \\ \cline{2-10}
\multicolumn{1}{|c|}{} & II & \multicolumn{1}{c|}{94.1\%} & 3.453 & \multicolumn{1}{c|}{94.7\%} & 2.419 & \multicolumn{1}{c|}{94.5\%} & 3.257 & \multicolumn{1}{c|}{95.1\%} & 2.806 \\ \cline{2-10}
\multicolumn{1}{|c|}{} & III & \multicolumn{1}{c|}{94.0\%} & 2.724 & \multicolumn{1}{c|}{94.2\%} & 1.883 & \multicolumn{1}{c|}{94.6\%} & 3.135 & \multicolumn{1}{c|}{95.5\%} & 2.935 \\ \hline
\end{tabular}
\label{tab3}
\end{table}

We again make some observations. First, across all settings, the proposed confidence interval achieves an empirical coverage probability close to or above the nominal level 95\%, confirming the validity of the proposed inference procedure. Second, when the number of subjects $K$ increases from 5 to 10, the confidence interval length becomes generally shorter, reflecting the efficiency gain from incorporating additional data sources. Finally, our inference method attains a consistent and robust performance for different heterogeneity levels and for both stable and unstable cases, with the coverage probability remaining close to the nominal level and interval length showing only modest variation.

\section{Approach-Avoidance Conflict Study via iEEG}
\label{sec:realdata}

We revisit the motivating example of the approach-avoidance conflict (AAC) study using intracranial EEG discussed in Section \ref{sec:introduction}. A key scientific objective is to uncover brain network patterns from the iEEG signals collected during the AAC task. Such an understanding of the cortical and limbic neural circuits that regulate approach and avoidance is crucial to understanding the neural underpinnings of both normal and excessive clinical anxiety. 

The data consists of $K=5$ subjects, each with iEEG signals collected from $p=6$ brain regions, including anterior cingulate cortex (ACC), dorsolateral prefrontal cortex (DLPFC), orbitofrontal cortex (OFC), amygdala, hippocampus, and insula. The iEEG signals were preprocessed and the theta waves were extracted, resulting in the observed trajectories with the number of time points $n$ ranging from 10,347 to 15,052, grouped into 130 to 190 trials, respectively, for the five subjects \citep{staveland2025circuit}. The data is publicly available at \url{https://zenodo.org/records/17726565}. There are also two stimulus signals measuring the risk and reward at a give time, which can be incorporated into our ODE system using the approach described in \citet[][Section 6]{dai2022kernel}. We first construct an estimate for each individual subject, then integrate them using the proposed method. Again, we compare it with the empirical risk minimization method and the invariance score method. We evaluate the performance of all methods in two ways. 

The first way of evaluation is out-of-sample generalization through leave-one-subject-out cross-validation. More specifically, we hold out the data of one subject entirely, apply the methods to the remaining subjects, and evaluate each method on the held-out subject at the trial level. For each held-out trial, we define the normalized prediction loss as,
\begin{equation}\label{eqn:normloss}
L(\widehat{F}) = \frac{ \sum_{j=1}^{p} \int_0^T \big\{ D_t X^{(k)}_j(t) - \widehat{F}_j(X^{(k)}(t), t) \big\}^2 \, dt }{ \sum_{j=1}^{p} \int_0^T \{D_t X^{(k)}_j(t)\}^2 \, dt },
\end{equation}
where the denominator measures the total variation of the derivative trajectory, and the nominator measures the prediction error. Here $L(\widehat{F}) \geq 0$, with a smaller value indicating a better predictive performance. Note that $L(\widehat{F})$ is a linear transformation of $R_S(\widehat{F})$ in \eqref{eqn:reward}, so ranking methods by the normalized prediction loss \eqref{eqn:normloss} is equivalent to ranking by the reward function $R_S(\widehat{F})$. We compare the three methods in a pairwise fashion: for each held-out trial, we record whether one method achieves a normalized loss no greater than that of another method i.e., $L(\widehat{F}_{\text{method 1}}) \leq L(\widehat{F}_{\text{method 2}})$, and report the frequency of such favorable comparisons across all trials. Table~\ref{tab:pairwise} reports the results. 

\begin{table}[t!]
\centering
\caption{Leave-one-subject-out cross-validation. Reported are the trial-level pairwise comparison frequencies, defined as the percentage of held-out trials for which the proposed method achieves a normalized loss \eqref{eqn:normloss} no greater than the competing method. A higher frequency indicates a more reliable generalization of the proposed method.}
\label{tab:pairwise}
\begin{tabular}{lrr} \hline
Subject ID & Proposed $vs$ ERM & Proposed $vs$ IVS \\ \hline
BJH016  & 124/190 (65.3\%) & 113/190 (59.5\%) \\
BJH021  & 92/164 (56.1\%) & 92/164 (56.1\%) \\
BJH025  & 85/132 (64.4\%) & 85/132 (64.4\%) \\
LL12    & 76/151 (50.3\%) & 59/151 (39.1\%) \\
SLCH002 & 64/130 (49.2\%) & 64/130 (49.2\%) \\ \hline
Overall & 441/767 (57.5\%) & 413/767 (53.8\%) \\ \hline
\end{tabular}
\end{table}

We see that the proposed method achieves a favorable comparison rate of 57.5\% against ERM, and 53.8\% against IVS across all held-out trials. At the subject level, the improvements are more evident for two subjects: BJH016 and BJH025, where the proposed method achieves favorable rates of 64.4-65.3\% against ERM, and 59.5-64.4\% against IVS. For subject LL12, the proposed method is comparable to ERM (50.3\%) but falls below IVS (39.1\%). An examination of the estimated weights reveals that this subject consistently receives substantially lower weight, with $\widehat{\omega}_{\text{stable},k} \approx 0.09$-$0.12$, while the other subjects receive the weights $\widehat{\omega}_{\text{stable},k} \approx 0.23$-$0.31$. This pattern suggests that this subject exhibits neural dynamics that are more distinct from the remaining subjects, and our robust weighting scheme down-weights this subject to help improve generalization to the majority of subjects.

\begin{figure}[t!]
\centering
\includegraphics[width=0.8\textwidth, height=5in]{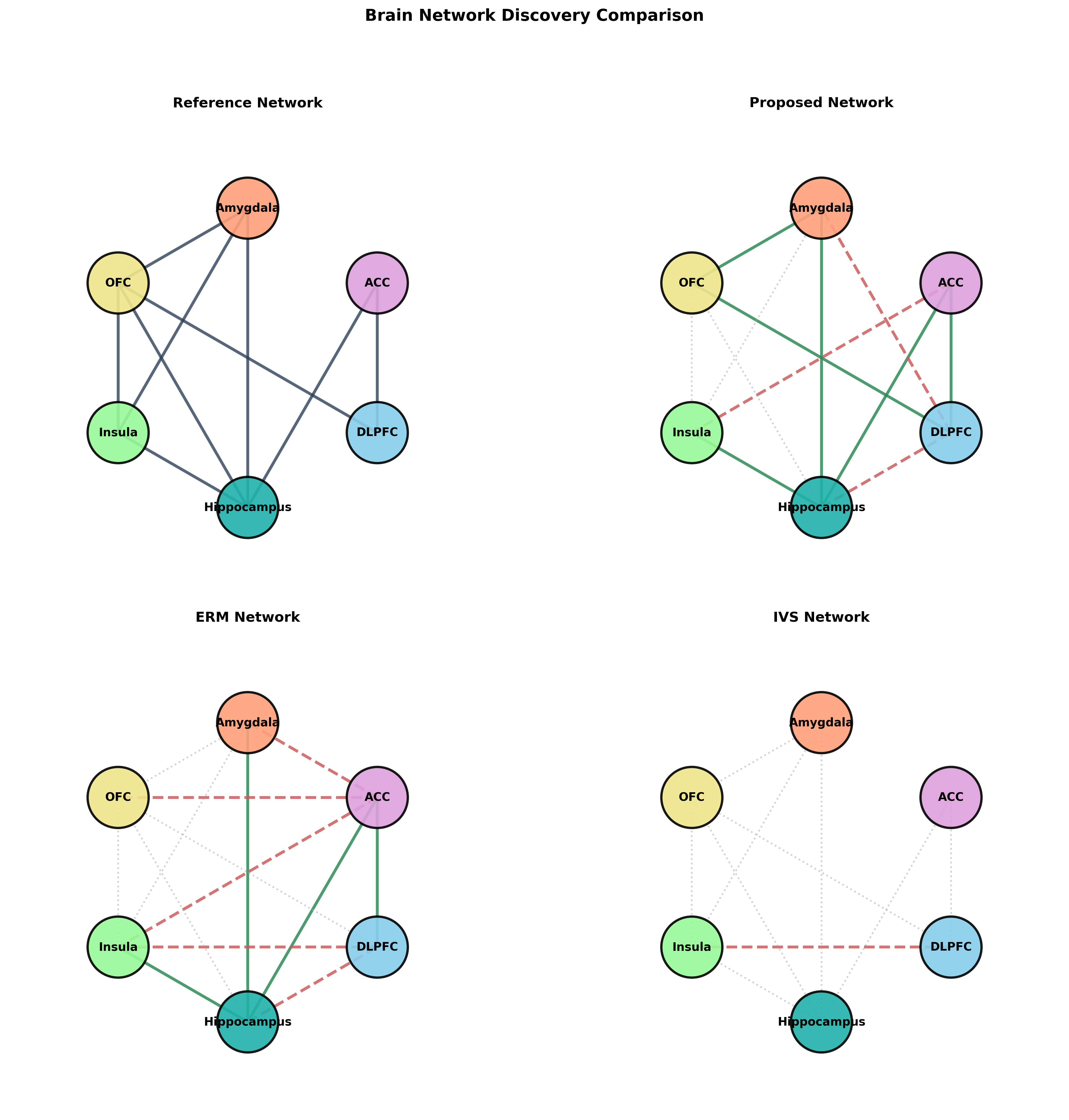}
\caption{Reconstruction of brain connectivity network during AAC by ERM (bottom left), IVS (bottom right), and the proposed method (top right), compared to the reference network (top left). Green solid edges denote the correctly recovered reference edges (true positive), red dashed edges denote the incorrectly recovered reference edges (false positive), and gray dotted edges denote the missed reference edges (false negative).}
\label{fig:brain_network_aac}
\end{figure}

The second way of evaluation is the reconstruction of brain connectivity network of the six brain regions. We apply the three methods using data from all five subjects. For the proposed method, we select the tolerance $d_n$ via the procedure as described in \eqref{eqn:dn} with $C_d = 0.01$, same as all simulations. The estimated stabilized weights $\widehat{\omega}_{\text{stable},1}, \ldots, \widehat{\omega}_{\text{stable},5}$ are $0.095, 0.229, 0.232, 0.209, 0.236$ for subjects LL12, BJH016, BJH021, BJH025, SLCH002, respectively. Notably, all five subjects receive nonzero weights, though LL12 is substantially down-weighted, consistent with the cross-validation findings above. Since there is no ground truth, we adopt the network from \citet{staveland2025circuit} as a reference network, and compare each method's reconstruction to this reference. Figure \ref{fig:brain_network_aac} reports the results.

If taking the reference network as the ground truth, the proposed method achieves an F1 score of 0.667, correctly recovering 6 of the 9 reference edges, while ERM achieves 0.444, with 4 of 9 edges, and IVS achieves 0.000, with 0 of 9 edges. Notably, the proposed method identifies several key connections in the prefrontal-hippocampal-amygdala circuit, including hippocampus-ACC, hippocampus-amygdala, hippocampus-insula, DLPFC-ACC, DLPFC-OFC, and amygdala-OFC. These connections are broadly consistent with both the reference AAC study of \citet{staveland2025circuit}, as well as the broader neuroscience literature on limbic-prefrontal circuits underlying approach-avoidance conflict. In particular, the recovered hippocampus-amygdala link aligns with well-established evidence that the hippocampus provides contextual information that interacts with amygdala threat evaluation during fear and anxiety processing \citep{ledoux2000emotion, phelps2004human}. The amygdala-OFC and DLPFC-OFC connections are also consistent with studies showing that the orbitofrontal cortex integrates affective signals from the amygdala with higher-level cognitive control and value evaluation from lateral prefrontal regions to guide decision making under uncertainty \citep{wallis2007orbitofrontal}. The recovered DLPFC-ACC link agrees with the widely studied prefrontal cognitive control network in which ACC monitors conflict and uncertainty while DLPFC implements top-down regulation and behavioral adjustment \citep{botvinick2004conflict, miller2001integrative}. In addition, the hippocampus-ACC interaction is consistent with evidence that hippocampal memory and contextual representations are integrated in medial prefrontal regions to guide adaptive behavior and emotional regulation \citep{fanselow2010functional, euston2012medial}. Finally, the hippocampus-insula connection is plausible in light of work suggesting that the insula integrates contextual memory with interoceptive and affective signals during uncertainty and anxiety \citep{craig2009how, paulus2006insular}. Together, these results provide evidence that the connectivity network recovered by our robust ODE approach captures scientifically meaningful neural circuitry involved in regulating approach-avoidance decisions.

In summary, the proposed robust learning framework yields more accurate brain network recovery and more generalizable dynamic model estimation compared to both ERM and IVS, demonstrating its practical usefulness for uncovering stable neural circuit patterns from heterogeneous multi-subject iEEG data.

\bibliographystyle{apalike} 
\bibliography{ref-ode}

\end{document}